\documentclass{optica-article}
\journal{opticajournal}
\articletype{Research Article}
\usepackage{amsmath,amsfonts}
\usepackage{algorithmic}
\usepackage{graphicx}
\usepackage{caption}
\usepackage{subcaption}
\usepackage{wrapfig}
\usepackage{textcomp}
\usepackage{xcolor}
\usepackage{stfloats}
\usepackage{url}
\usepackage{textcomp}
\usepackage{siunitx}

\begin{document}
\title{Accumulation of cross-channel non-linear interference in dispersion-managed and disaggregated optical network segments}
\author{Elliot London\authormark{*}, Emanuele Virgillito, Andrea D'Amico, and Vittorio Curri}
\address{Department of Electronics and Telecommunications, Politecnico di Torino, Corso Duca degli Abruzzi 24, 10129 Torino, Italy}
\email{\authormark{*}elliot.london@polito.it}
\begin{abstract}
We evaluate the generation of the cross-channel interference (XCI) for coherent transmission through a variety of dispersion-managed segments in a disaggregated optical network framework, using split-step Fourier method (SSFM) simulations and an implementation of the Gaussian noise (GN) model.
We observe that the small inline residual dispersion remaining after each span affects the accumulation of the XCI, causing GN model predictions to no longer be conservative.
We find an asymptotic upper bound to this additional accumulation, providing a worst-case prediction, and observe that this depends upon the residual dispersion within the link.
This upper bound scales similarly to the self-channel interference (SCI) accumulation, and is well characterized by the parameters of the underlying fiber spans and the transmitted signals.
\end{abstract}
\section{Introduction}
\label{sec:introduction}
As network operators strive to provide data throughput levels that satisfy ever-growing user demands~\cite{cisco2019}, the maximization of network capacity has become an increasingly relevant goal~\cite{bayvel2016maximizing}.
Besides the installation of new fiber spans, which is often a prohibitively expensive undertaking~\cite{wellbrock2014will}, ideas are being formulated to increase capacity and throughput by better utilizing existing network infrastructures.
Two major examples in this direction are the increasing number of projects and technological innovations that aim to enable wideband and disaggregated network infrastructures~\cite{kozdrowski2020ultra,riccardi2018operator,fischer2018maximizing,paolucci2020telemetry,casellas2018control}, with the former attempting transmission in currently-unused transmission bandwidths, and the latter enabling open and multi-vendor operation, increasing network flexibility and supporting software-defined networking (SDN) approaches~\cite{curri2020software}.
These two approaches are not mutually exclusive; in fact, upgrades towards wideband infrastructures are anticipated to be performed progressively~\cite{ahmed2021c+,paolucci2020disaggregated,moniz2020design}, where network segments may have varying transmission bandwidths, a framework which would correspondingly require a disaggregated approach to quality of transmission (QoT) estimation.

Concerning these progressive network upgrades, most backbone networks transmit dual-polarization (DP) coherent signals, supporting data transmission with a variety of modulation formats; an example being the next-generation 400G-ZR+ implementation, which enables 64\,GBd transmission in a 75\,GHz, wavelength division multiplexing (WDM) grid, for total bandwidths of up to 4.8\,THz within the C-band~\cite{pincemin2022927,pincemin2022end}.
On the other hand, many metro and access networks transmit signals using intensity-modulated direct-detected (IMDD) transceivers that deliver up to 10\,Gb/s through dispersion managed optical line systems (OLS). 
This requires optical chromatic dispersion compensation inline, using dispersion compensation units (DCUs), placed at the end of each (or a subset of) fiber spans, within the fiber infrastructure.
These network segments are also experiencing progressive upgrades towards coherent technologies.
In some cases this upgrade may still be too costly, creating scenarios where it would be useful to propagate IMDD and coherent signals alongside each other in dispersion-managed network segments~\cite{virgillito2021qot}.
Consequently, as these upgrades are performed, the problem of how to estimate the QoT of coherent signals through dispersion-managed links remains relevant.

\begin{figure}[b]
    \centering
    \includegraphics[width=0.75\linewidth]{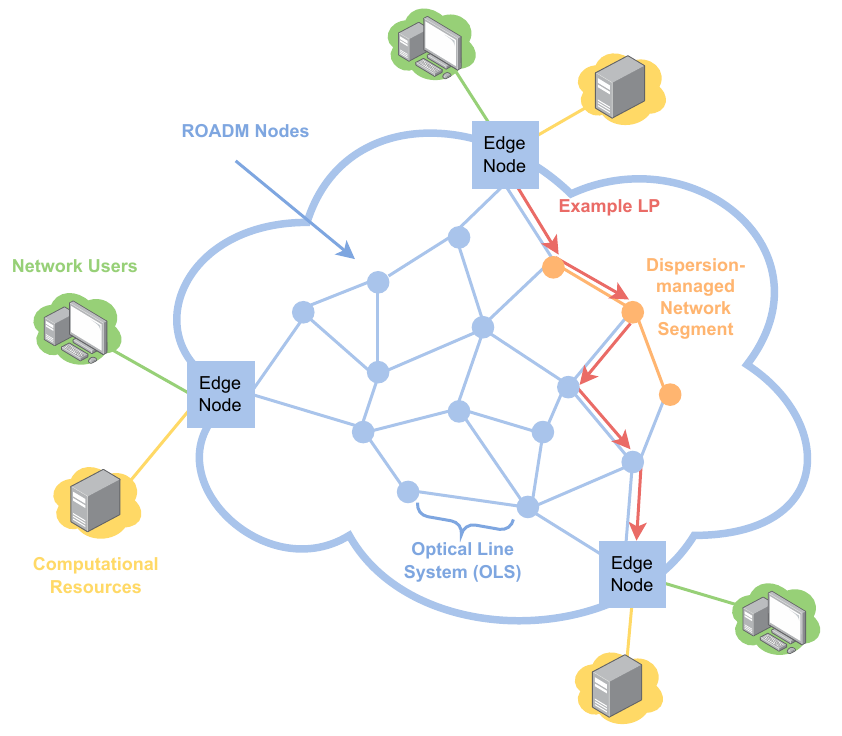}
    \caption{An example disaggregated optical network infrastructure. Data is transmitted along LPs that pass through successive OLSs from a given source to destination. Within our investigation, some links may still depend upon in-line dispersion management in the form of DCUs, highlighted here in orange.}
    \label{fig:network}
\end{figure}

Disaggregated networks are enabled by the inclusion of open and disaggregated reconfigurable optical add-drop multiplexers (ROADMs) at the termination of their constituent OLSs, in turn enabling routing of open and disaggregated lightpaths (LPs).
An example of a disaggregated optical network infrastructure is shown in Fig.~\ref{fig:network}; an LP may originate and terminate at any given node.
Given a pair of source-destination nodes, the path over which the signal will be routed is determined primarily through the estimation of the available QoT on each path, i.e. performing a path feasibility assessment.
A QoT degradation occurs as the signal propagates through the network links, with both linear and nonlinear contributions.

Linear impairments primarily are due to amplified spontaneous emission (ASE) noise, which arises from the in-line amplifiers (ILAs).
Nonlinear impairments are exclusively due to the nonlinear interference (NLI) noise, arising from interactions between the signal and the transmission medium (the fiber spans).
The NLI has two significant contributors; the self-channel interference (SCI) and cross-channel interference (XCI)~\cite{poggiolini2012gn}, generated by the channel under test (CUT) upon itself, and by the interfering channels on the CUT, respectively~\cite{carena2014egn}.
All other NLI contributions are typically grouped together as four-wave mixing (FWM) effects, and are negligible for most realistic operating scenarios~\cite{dar2013properties}.

When working within a disaggregated framework it is necessary to take an equally disaggregated standpoint when evaluating QoT degradation.
In uncompensated transmission (UT) this is required due to the coherent accumulation of the SCI contributor~\cite{d2020quality}, which causes the SCI generation for a given span to depend upon previously traversed fiber spans, for a given optical channel.
%
%
On the other hand, the XCI accumulates incoherently, and by quantifying the upper bound of the SCI generation, it is possible to create a fully disaggregated NLI model~\cite{london2020simulative,d2020quality}.
For disaggregated networks which feature dispersion management, the presence of DCUs has been observed to impact the accumulation of the SCI, worsening the coherent accumulation effect, and requiring NLI model revisions to ensure accuracy~\cite{virgillito2022spatially}.
The effects of XCI generation within this scenario has not yet been characterised from a disaggregated standpoint, and is an essential step to ensure that NLI models remain accurate for coherent transmission within dispersion-managed network segments.

Within this work we consider a disaggregated network segment consisting of two distinct, periodic, and dispersion-managed OLSs, composed of 10 and 20 fiber spans, respectively, with each span followed by an ILA and DCU.
The accumulation of XCI through these segments is then analyzed using split-step Fourier method (SSFM) simulations to quantify the nonlinear impairment of a variety of pump-and-probe scenarios, for a wide range of transmission configurations and characteristic fiber parameters.
The DCUs compensate for the accumulated dispersion of the fiber, but leave a controlled amount of residual dispersion, $D_{\mathrm{RES}}$, after each fiber span; this quantity is varied and the effect upon the XCI generation is observed, forming the core focus of this investigation.
These results are compared to an implementation of the Gaussian noise (GN) model using the open-source GNPy library~\cite{ferrari2020gnpy,gnpy-github}, providing a reference point for XCI accumulation in a scenario without dispersion management.

The rest of this paper is divided into the following sections; in Sec.~\ref{sec:architectures} the optical network architectures under investigation within this work are described, along with the disaggregated model used for QoT analysis.
In Sec.~\ref{sec:simulations} we introduce the simulation strategy used to calculate the overall XCI generated during propagation through the OLS elements.
In Sec.~\ref{sec:results} the results of the simulations are presented and discussed, and the paper is concluded in Sec.~\ref{sec:conclusion}.
\section{Dispersion-Managed and Disaggregated Architectures}
\label{sec:architectures}
\begin{figure}[b]
    \centering
    \includegraphics[width=0.95\linewidth]{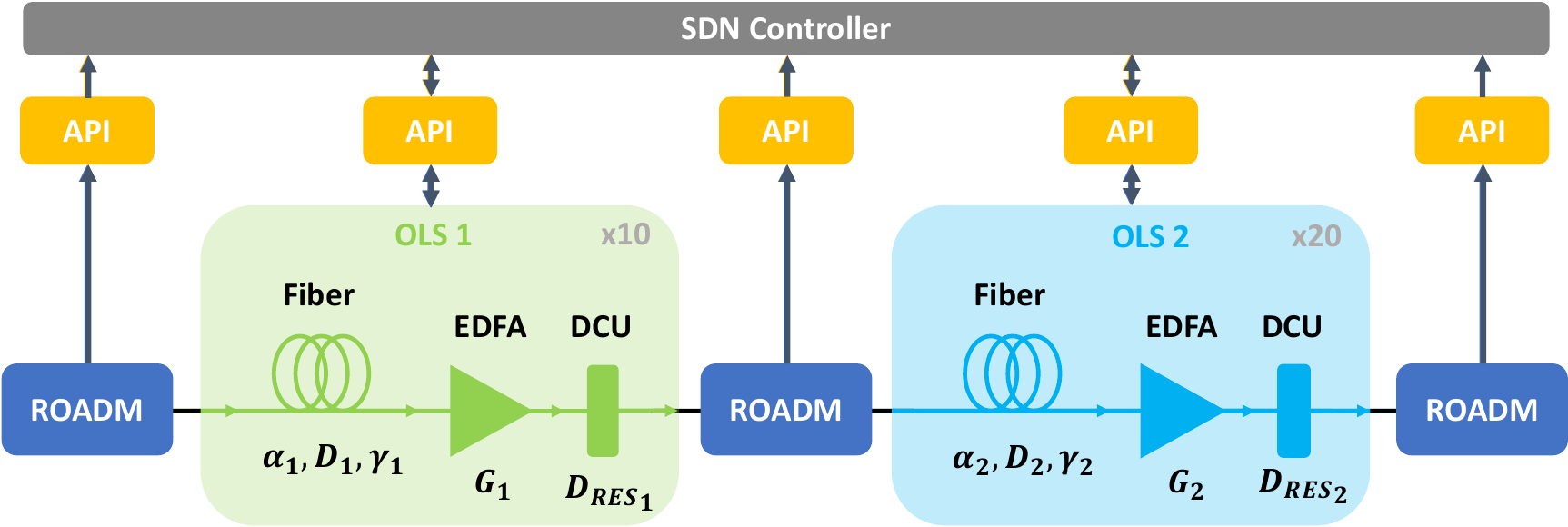}
    \caption{An outline of a dispersion-managed segment of a disaggregated optical network, consisting of OLSs with two distinct parameters, managed by a SDN controller with use of multiple APIs.}
    \label{fig:segment}
\end{figure}

A schematic of the basic structure of the disaggregated network segment featuring dispersion management that is considered in this work is given in Fig.~\ref{fig:segment}. 
Here, OLSs with two distinct configurations may be managed by SDN controller(s) through the use of multiple application programming interfaces (APIs).
In this framework, a LP is assigned between a source and destination, and the signal is transmitted from the source, beginning propagation from a node that includes a disaggregated ROADM.
As mentioned within the previous section, the history of the LP under test affects the accumulation of the SCI, an effect which has already been investigated within~\cite{london2022modelling,virgillito2022spatially} and is not the focus of this work.
Bearing this in mind, we consider the first ROADM node to correspond to the LP source, and the first element within the link to be OLS1.

Next, the signal passes through the first OLS segment (OLS1).
Each fiber span is characterized most significantly by its attenuation, $\alpha_1$\,[dB/km], chromatic dispersion, $D_1$, in ps\,/\,(nm$\cdot$km), and nonlinear coefficient, $\gamma_1$.
Each span is  followed by an erbium-doped fiber amplifier (EDFA) with a gain value, $G_1$, that fully recovers the fiber loss (operating in transparency), which is a condition that may be lifted without any loss of generality.
Following the amplifier, a DCU module is present, compensating for fiber dispersion and producing a residual dispersion, $D_{\mathrm{RES},1}$, given in general for any OLS:
\begin{equation}
    D_\mathrm{RES} = D L_s + D_\mathrm{DCU}\;,
\end{equation}
where $D$ is the dispersion of the fiber span, and $L_s$ is the fiber span length. 
In this campaign $D_{\mathrm{RES}}$ values of 40, 80, and 160\,ps\,/\,nm were selected for investigation, which span a typical range of residual dispersion values that are encountered when DCUs are used~\cite{beygi2014coded}.
We remark that these DCUs are considered to only apply chromatic dispersion, functioning as linear elements which do not produce any XCI when the signal propagates through them.

After this, the signal then passes successively into 10 spans with the same format and line elements, and each with identical physical parameters.
The signal then passes through another ROADM node, and into OLS2, made up of 20 different successive sections, all characterized by values of $\alpha_2$, $D_2$, $\gamma_2$ and $G_2$. 
After traversing these spans, the signal reaches the destination node represented by a final ROADM and is received.
A total of 10 spans were chosen for OLS1 and 20 spans for OLS2, ensuring that behavior of the XCI accumulation in the presence of residual dispersion is fully characterized.
The choice of 10 spans for OLS1 and 20 spans for OLS2 is motivated solely to best observe the behavior of the XCI accumulation, such that any macroscopic trends are fully visible, both before and after the change of dispersion within the segment.

The QoT impairment of a given LP with a wavelength, $\lambda$, is most commonly quantified using the generalized signal-to-noise ratio (GSNR) by making the assumption that the LP may be modelled as an additive and white Gaussian noise (AWGN) channel~\cite{carena2013impact,curri2020software}, an approach which enables use of the well-verified and accurate GN model family of modelling systems for uncompensated transmission~\cite{poggiolini2016recent,johannisson2013perturbation,serena2013accuracy}.
The GSNR for a single wavelength is defined as:
\begin{equation}
    \mathrm{GSNR}_{\lambda} = \left(\mathrm{OSNR}_{\lambda}^{-1} + \mathrm{SNR}_{\mathrm{NL};\lambda}^{-1}\right)^{-1}\;,
\end{equation}
where the two contributors to the GSNR are the OSNR and SNR$_\mathrm{NL}$; the former provides all linear contributions, which is primarily the amplified spontaneous emission (ASE) noise that arises from the amplification process, and the latter includes all nonlinear contributions, which is primarily the NLI noise in the case of standard C-band transmission scenarios.

Within this work we wish to investigate XCI generation within a disaggregated network region that contains multiple dispersion-managed OLSs, for coherent LPs. 
This represents a worst-case scenario for a network undergoing upgrades from pure IMDD to fully coherent transmission.
Consequently, we focus our investigation solely upon SNR$_{\mathrm{NL}}$, neglecting linear impairments including ROADM filtering penalties and amplifier noise.
For a scenario with multiple OLSs, containing a total of $N_s$ successive fiber spans, the total nonlinear impairment, SNR$_{\mathrm{NL};\mathrm{tot}}$ for a given wavelength is given by~\cite{london2022modelling}:
\begin{equation}
    \mathrm{SNR}_{\mathrm{NL};\mathrm{tot}} = \left(\sum^{N_s}_{i=1} \frac{(1 + C_\infty)P_{\mathrm{SCI};\lambda} + P_{\mathrm{XCI};\lambda}}{P_{\lambda}}\right)^{-1}\;,
\end{equation}
where $P_{\mathrm{SCI};\lambda}$ and $P_{\mathrm{XCI};\lambda}$ are the SCI and XCI powers, respectively, and C$_\infty$ is a quantity known as the SCI coherency coefficient~\cite{d2020quality}.
We remark that C$_\infty$ is well-defined for uncompensated transmission scenarios, but defining this parameter in dispersion-managed scenarios has been thoroughly investigated in~\cite{virgillito2022spatially}.
In any case, we are solely interested in the generation of XCI from a spatially disaggregated standpoint, and as such we consider only the XCI contribution to the NLI, which is generated only by interfering pumps upon the CUT.
The total XCI QoT degradation is given by:
\begin{equation}
    \mathrm{SNR}_{\mathrm{XCI;tot}} = \left(\sum^{N_s}_{i=1}\frac{P_{\mathrm{XCI};\lambda}}{P_{\lambda}}\right)^{-1}\;,
    \label{eq:snr-xci}
\end{equation}
by calculating this quantity after each span within the network configuration it therefore becomes possible to investigate the accumulation of the XCI.
\section{Simulation Framework}
\label{sec:simulations}
The SNR of the XCI may be estimated through the use of accurate simulation tools, specifically in this work we have made use of our SSFM framework written in the MATLAB\textsuperscript{\textregistered} programming environment, derived from an implementation described in detail in~\cite{pilori2017ffss}.
This tool allows the XCI to be estimated on a per-span basis for a wide variety of dispersion-managed OLSs and transmission configurations.
The XCI may be calculated by performing pump-and-probe simulations, where the NLI noise generated by a single interfering channel is calculated by a single CUT with a sufficiently low input power; in this way, the SCI contribution is negligible, leaving only the XCI arising from the interfering channel~\cite{thiele2002investigation,london2020simulative}.

\begin{figure}[b]
    \captionsetup{singlelinecheck=false, font=small, justification=raggedright}
    \centering
    \begin{subfigure}[t]{0.49\linewidth}
        \includegraphics[width=1\linewidth]{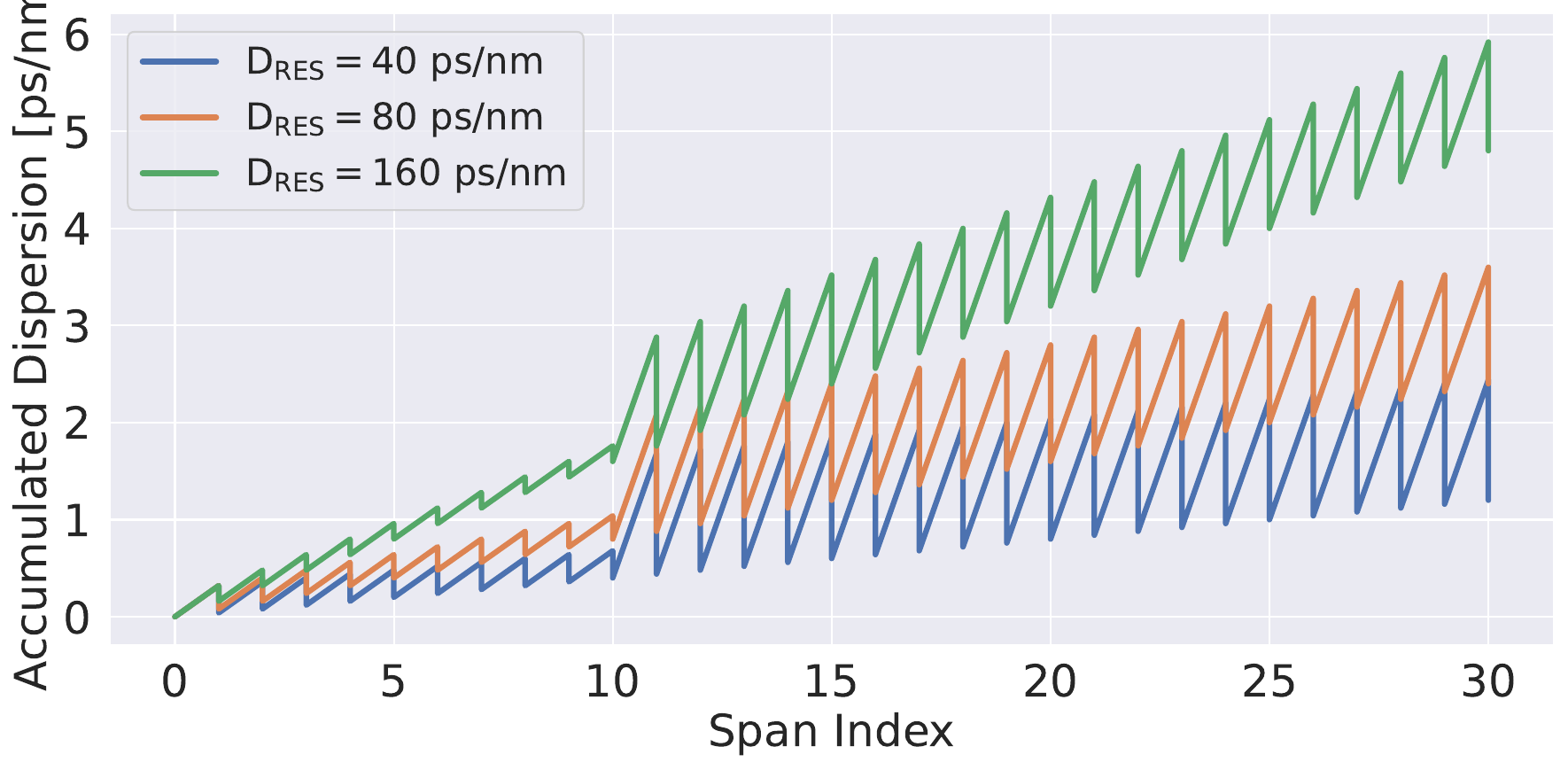}
        \footnotesize
        \vspace{-6mm}
        \caption{}
        \label{fig:osac_3_dispmap_d4_d16}
    \end{subfigure}
    \begin{subfigure}[t]{0.49\linewidth}
        \includegraphics[width=1\linewidth]{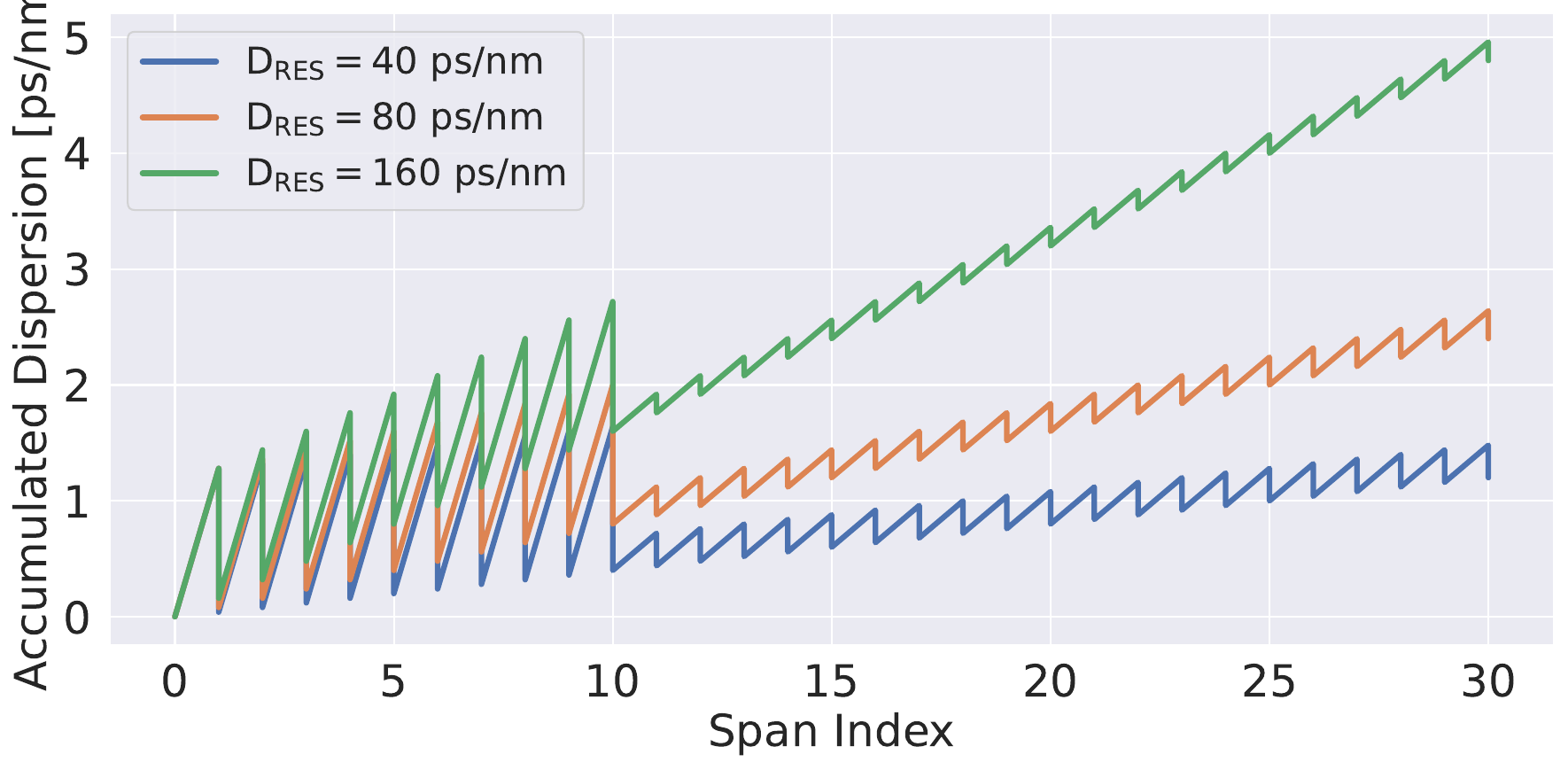}
        \footnotesize
        \vspace{-6mm}
        \caption{}
        \label{fig:osac_3_dispmap_d16_d4}
    \end{subfigure}
    \caption{The accumulated dispersion verses the span index for the entire disaggregated network segment, for the three considered $D_RES$ values of 40, 80, and 160\,ps\,/\,(nm$\cdot$km), for (a): $D_1 = 4$\,ps\,/\,(nm$\cdot$km) and $D_2 = 16$\,ps\,/\,(nm$\cdot$km), and (b): $D_1 = 16$\,ps\,/\,(nm$\cdot$km) and $D_2 = 4$\,ps\,/\,(nm$\cdot$km).}
    \label{fig:dispersion_maps}
\end{figure}

We begin by defining the line configuration: as we wish to investigate a realistic, worst-case scenario within a disaggregated optical network, we consider what we denote as a disaggregated network segment, highlighted in orange in Fig.~\ref{fig:network}, shown in Fig.~\ref{fig:segment} and explained in detail in Sec.~\ref{sec:architectures}.
The majority of the fiber parameters are set to be equal throughout the simulation campaign, specifically the nonlinear coefficients, $\gamma=1.27$\,1/W/km, and loss values, $\alpha=0.2$\,dB\,/\,km.
For this campaign, we focus upon changing solely the dispersion values of the two OLSs, $D_1$ and $D_2$, and the uniform level of residual dispersion produced by each DCU.
Two distinct dispersion values have been considered: $D_1 = 4$\,ps\,/\,(nm$\cdot$km) and $D_2 = 16$\,ps\,/\,(nm$\cdot$km), and vice versa, creating one scenario where the dispersion coefficient of the fiber spans increases as the signal passes into OLS2, and one where it decreases.
These two configurations generate two corresponding dispersion maps, which are shown in Figs.~\ref{fig:osac_3_dispmap_d4_d16} and~\ref{fig:osac_3_dispmap_d16_d4}, for the $D_1 = 4$\,ps\,/\,(nm$\cdot$km) and $D_2 = 16$\,ps\,/\,(nm$\cdot$km) scenario, and the $D_1 = 16$\,ps\,/\,(nm$\cdot$km) and $D_2 = 4$\,ps\,/\,(nm$\cdot$km) scenario, respectively.
Additionally we have considered two different values for fiber lengths, $L_s$, of 80 and 50\,km, with all fibers in both OLSs set so that the fiber lengths are identical.

Concerning the spectral load, we have considered a single CUT (the probe) and a single interfering channel (the pump) within a uniform WDM grid, with the CUT located at the center of the C-band (193.9\,THz).
Both channels are transmitted with a DP-16-QAM modulation format and a large amount (102400\,ps\,/\,nm) of predistortion, permitting the signal to be considered as Gaussian distributed prior to transmission, and excluding any effects due to signal gaussianization~\cite{poggiolini2014simple, virgillito2022spatially}, thus allowing the coherent accumulation effect to be isolated.
Baud rates, $R_s$, of 32 and 64\,GBaud have been considered, on the WDM fixed grid of 37.5 and 75\,GHz, respectively.
Two spectral separation values between the CUT and interfering channel have been considered, corresponding to multiples of two and four of the WDM grid width, i.e. 75 and 150\,GHz for the 32\,GBaud scenario, and 150 and 300\,GHz for the 64\,GBaud scenario. 
This approach is equivalent to isolating the $i=2$ and $i=4$ pumps in the XCI contributions of Eq.~\ref{eq:snr-xci}.
The power of the CUT and interfering channel are set to -20 and 1\,dBm, respectively, to ensure that negligible SCI is generated.
For both channels independent pseudo-random binary sequences (PRBSs) are used to generate the signal, for each polarization, using a 17th degree polynomial.

For each line and spectral configuration the signal is propagated through the system and received by a coherent digital signal processing (DSP) module.
The XCI evolution during propagation is able to be captured by our simulation tool by placing a receiver at the end of each span to capture the SNR of the XCI.
The DSP receiver is made up of a resampling filter that isolates the CUT, followed by an ideal analog-to-digital converter (ADC), then followed by a chromatic dispersion compensation (CDC) unit, which compensates for $D_\mathrm{RES}$ that is introduced after each span.
After this, the signal passes into an adaptive equalizer module that makes use of a least-mean squares (LMS) algorithm, with the number of taps set to 42, and an adaptation coefficient, $\mu=10^{-4}$, followed by a carrier phase estimation (CPE) module, which completely recovers the phase noise.
Although not realistic, such large number of taps permits a large back-to-back SNR and, allowing even small amounts of XCI noise to be observed.
The SNR of the XCI contribution is then evaluated by estimating the error vector magnitude (EVM) of the received constellation at 1 sample per symbol after the CPE, at the end of each fiber span.

\section{Results and Analysis}
\label{sec:results}
\begin{figure}[b]
    \captionsetup{singlelinecheck=false, font=small, justification=raggedright}
    \centering
    \begin{subfigure}[t]{0.49\linewidth}
        \includegraphics[width=1\linewidth]{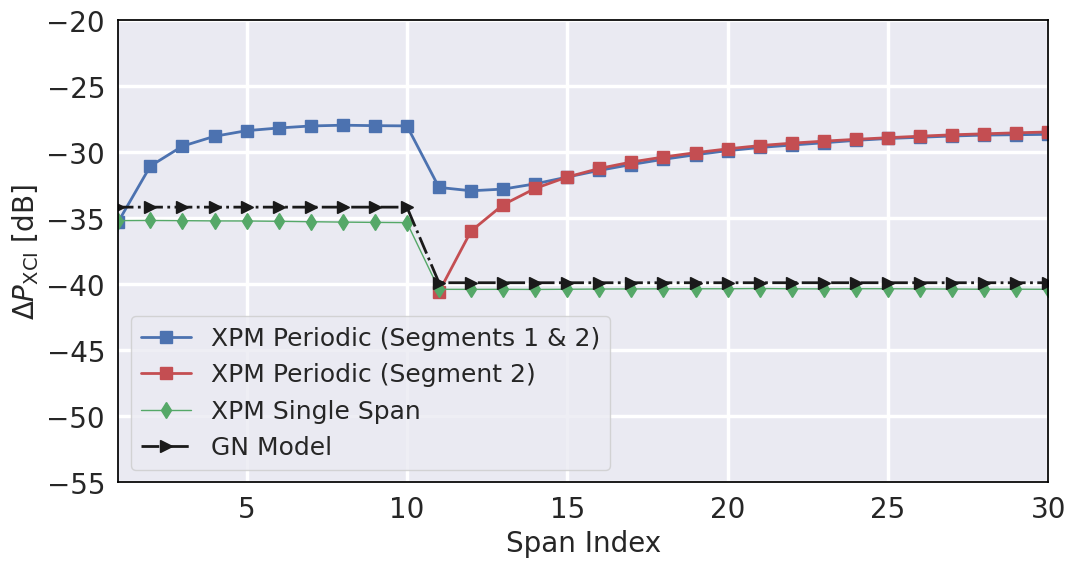}
        \footnotesize
        \vspace{-10mm}
        \caption{}
        \label{rs32_df37_d04_d16_pump02_res0040_loss02}
    \end{subfigure}
    \begin{subfigure}[t]{0.49\linewidth}
        \includegraphics[width=1\linewidth]{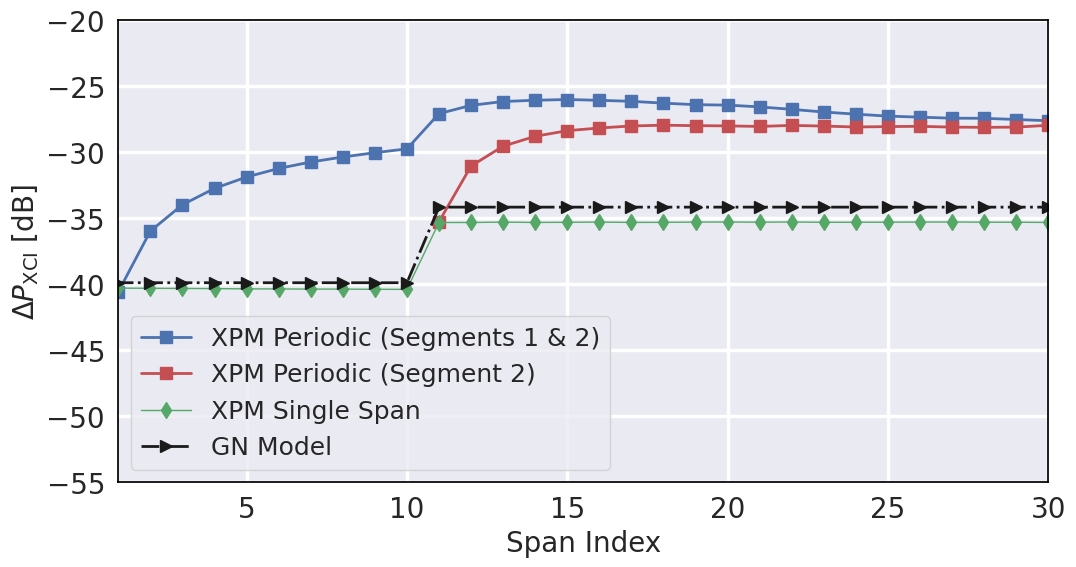}
        \footnotesize
        \vspace{-10mm}
        \caption{}
        \label{rs32_df37_d16_d04_pump02_res0040_loss02}
    \end{subfigure}
    \begin{subfigure}[t]{0.49\linewidth}
        \includegraphics[width=1\linewidth]{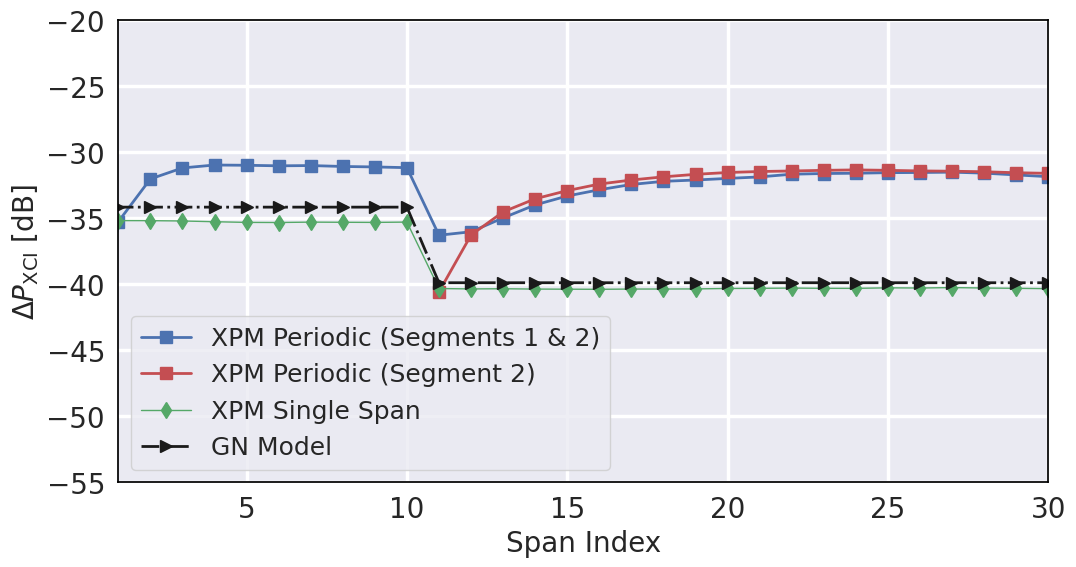}
        \footnotesize
        \vspace{-10mm}
        \caption{}
        \label{rs32_df37_d04_d16_pump02_res0080_loss02}
    \end{subfigure}
    \begin{subfigure}[t]{0.49\linewidth}
        \includegraphics[width=1\linewidth]{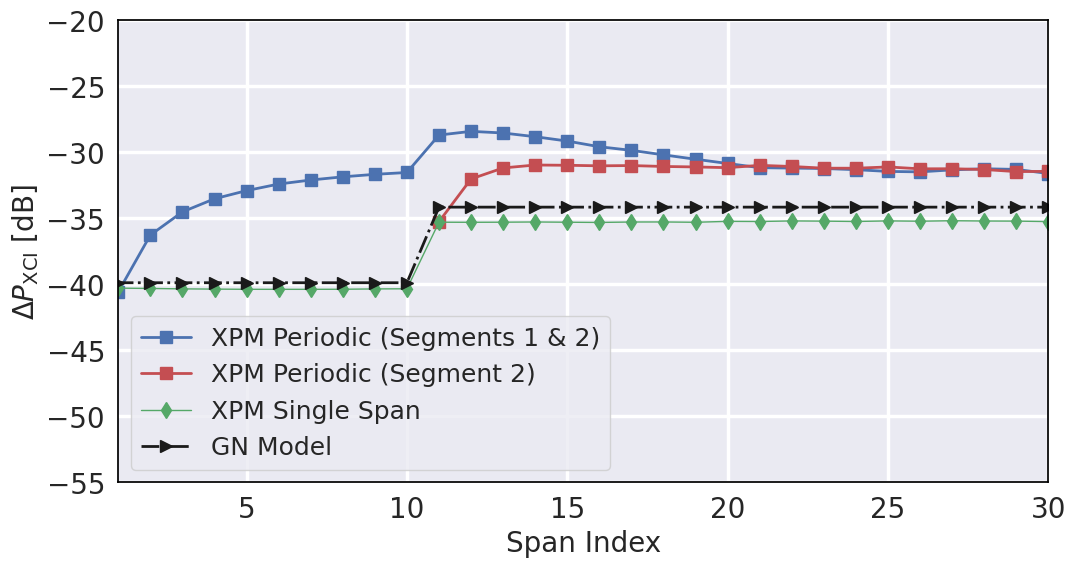}
        \footnotesize
        \vspace{-10mm}
        \caption{}
        \label{rs32_df37_d16_d04_pump02_res0080_loss02}
    \end{subfigure}
    \begin{subfigure}[t]{0.49\linewidth}
        \includegraphics[width=1\linewidth]{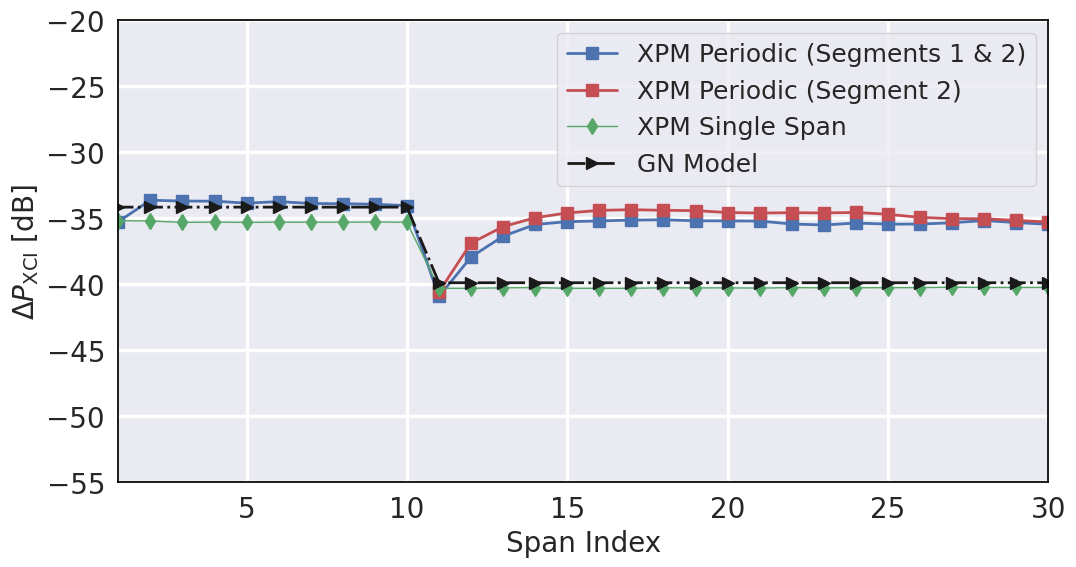}
        \footnotesize
        \vspace{-10mm}
        \caption{}
        \label{rs32_df37_d04_d16_pump02_res0160_loss02}
    \end{subfigure}
    \begin{subfigure}[t]{0.49\linewidth}
        \includegraphics[width=1\linewidth]{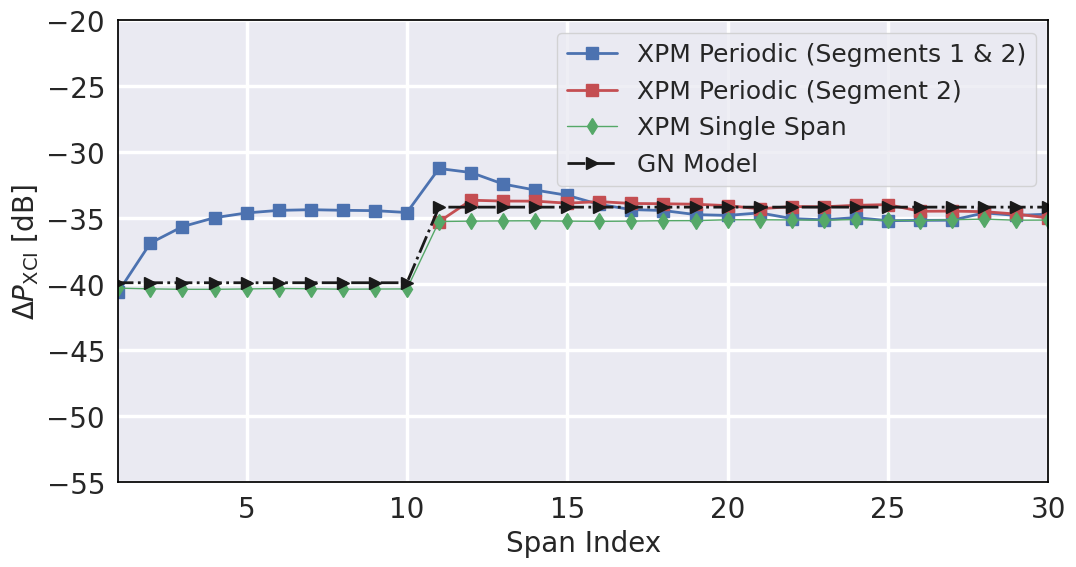}
        \footnotesize
        \vspace{-10mm}
        \caption{}
        \label{rs32_df37_d16_d04_pump02_res0160_loss02}
    \end{subfigure}
    \caption{The NLI accumulation for a disaggregated network segment, given in terms of XCI power gradient versus span index. Included are the simulation results for the entire segment (blue lines), for the final 20 spans (red lines), each span evaluated independently (green lines) and an implementation of the GN model (black dashed lines). The six configurations are as follows; (a): $D_1=4$, $D_{\mathrm{RES}}=40$, (b): $D_1=16$, $D_{\mathrm{RES}}=40$, (c): $D_1=4$, $D_{\mathrm{RES}}=80$, (d): $D_1=16$, $D_{\mathrm{RES}}=80$, (e): $D_1=4$, $D_{\mathrm{RES}}=160$, (f): $D_1=16$, $D_{\mathrm{RES}}=160$ in units of ps\,/\,(nm$\cdot$km) and ps\,/\,nm, respectively.}
    \label{fig:pump2_results}
\end{figure}

We present a selection of results from the simulation campaign in Fig.~\ref{fig:pump2_results}, for a single interfering pump spaced 75\,GHz away from the CUT, with a baud rate of $R_s$ = 32\,GBaud, and D$_\mathrm{RES}$ values of 40, 80, and 160\,ps\,/\,nm.
We give these results in terms of XCI power gradient, $\Delta P_{\mathrm{XCI}}$ (in dBm) against the index of the investigated fiber span.
For a fiber span index $i$, the XCI power gradient is defined as:
\begin{equation}
    \Delta P_{\mathrm{XCI},i} = P_{\mathrm{XCI},i} - P_{\mathrm{XCI},i-1}\,,
\end{equation}
where $P_{\mathrm{XCI},i}$ is the accumulated XCI power calculated after the DSP receiver placed at the termination of the $i$th fiber span.
Within these plots we present four different methods of calculating the XCI: in blue and red the reference gradient curves are shown, where the disaggregated segment is considered in its entirety, starting from the inputs of OLS1 and OLS2, respectively.
These scenarios show the aggregated effect of the XCI, evolving span-by-span, along with any additional impairments that arise due to inline dispersion compensation.
In black the results of the incoherent GN (IGN) model calculated using GNPy are given, where each span contribution is considered to be independent of all other fiber spans.
In green, the results are shown for a SSFM campaign where the XCI is calculated for each span, independently of all other spans.
These values are obtained by switching on the Kerr effect, (enabling the NLI), one span at a time, which is implemented in the SSFM engine by setting $\gamma=0$ for all spans except the $i$th, providing the intrinsic XCI noise contribution.
Comparing the black and green curves, where the XCI is fully incoherent, we observe that the IGN model provides an accurate, slightly conservative estimation of the intrinsic XCI noise power.

On the other hand, considering first the periodic cases where $D_1=4$\,ps\,/\,(nm$\cdot$km) (Figs.~\ref{rs32_df37_d04_d16_pump02_res0040_loss02}, \ref{rs32_df37_d04_d16_pump02_res0080_loss02}, \ref{rs32_df37_d04_d16_pump02_res0160_loss02}), in the first 10 spans the XCI has larger $\Delta P_{\mathrm{XCI}}$ values with respect to the IGN predictions and pure XCI simulations, eventually reaching a stable equilibrium for all cases.
Consequently, the total XCI introduced by each span cannot be estimated with the IGN model, which underestimates it by several dB and is no longer conservative.
We observe that, as $D_{\mathrm{RES}}$ increases, this discrepancy decreases, and for the 160\,ps\,/\,nm case in Fig.~\ref{rs32_df37_d04_d16_pump02_res0160_loss02}, after a single span the XCI has reached an asymptotic level which corresponds to the GN model.

Moving to the final 20 spans in these $D_1 = 4$ ps/nm/km scenarios, the $\Delta P_{\mathrm{XCI}}$ value in the periodic scenario starting from OLS1 (blue curve) drops as the signal passes into OLS2, before slowly increasing once more towards the same asymptote encountered within OLS1.
For the periodic scenario starting from OLS2 (red curve) a similar behavior is seen, except for a differing accumulation within the first few fiber spans.
This reveals two important points: firstly, the residual dispersion induces a memory effect into the XCI accumulation, no longer implying that each span may be considered independently of the previously crossed fiber spans.
This also implies that there may be a significant underestimation of the XCI if the IGN model is used for a dispersion-managed segment with a sufficiently low residual dispersion.
Secondly, the asymptote of the XCI accumulation depends only upon the parameters of the fiber spans located within the first OLS.

Next, considering the periodic cases where $D_1=16$\,ps\,/\,(nm$\cdot$km) (Figs.~\ref{rs32_df37_d16_d04_pump02_res0040_loss02}, \ref{rs32_df37_d16_d04_pump02_res0080_loss02}, \ref{rs32_df37_d16_d04_pump02_res0160_loss02}), a similar but opposite behavior is observed; initially, the XCI accumulates towards an asymptotic value, and $\Delta P_{\mathrm{XCI}}$ jumps upwards when the signal passes into OLS2, then decreasing towards the OLS1 asymptote.
Much like the $D_1=4$\,ps\,/\,(nm$\cdot$km) cases, increasing the residual dispersion within the system reduces the discrepancy between the observed XCI accumulation and the IGN model.

We also note that the investigations which have been performed for fiber lengths of 50\,km and baud rates of 64\,GBaud, yielded results with identical behaviors to those observed within Fig.~\ref{fig:pump2_results}, with a simple shift in SNR values.
As a specific example, for a fiber length of 50\,km, this shift corresponded to the change in XCI expected due to the change in effective length, $L_{\mathrm{eff}}$, and for the other parameters represented a shift in $P_{\mathrm{XCI}}$.
Additional minor investigations were made for $D_\mathrm{RES}$ larger than 160\,ps\,/\,nm and smaller than 40\,ps\,/\,nm: for values of $D_\mathrm{RES}$ significantly larger than 160\,ps\,/\,nm the XCI accumulation very quickly falls/rises to the level given by the IGN model as a result of the residual dispersion being large enough that no memory is observed between successive fiber spans.
For $D_\mathrm{RES}$ values smaller than 40\,ps\,/\,nm, the number of spans required to reach the asymptotic level progressively rose to an unrealistic level, with the accumulation increasing for dozens of spans before reaching a stable level.

Combining all of this information and the behavior presented in Figs.~\ref{rs32_df37_d04_d16_pump02_res0040_loss02}~--~\ref{rs32_df37_d16_d04_pump02_res0160_loss02}, we observe that the accumulation of XCI depends upon the parameters of the first fiber span type within the network segment and the residual dispersion present within the segment.
This behavior may be fully characterized using a parameter, $C_{ij}$, that provides the amount of additional XCI accumulation generated by propagation through each fiber span, following the approach outlined in~\cite{virgillito2022spatially}.
The XCI power gradient for a given span, $\Delta P_{\mathrm{XCI,i}}$, may be written in terms of $C_{ij}$ as:
\begin{equation}
    \Delta P_{\mathrm{XCI,i}} = \sigma_i^2 + 2\sum^{i-1}_{j=1} C_{ij} \sigma_i \sigma_j\;,
    \label{eq:cij}
\end{equation}
where $\sigma_i$ is the so-called "pure" XCI power generated at the $i$th fiber span, given by the green lines in Fig.~\ref{fig:pump2_results} and not counting any additional coherent impairments induced by residual dispersion present within the line.
The presence of residual dispersion causes the XCI contributions between two spans $i$ and $j$ to sum coherently at the receiver after CDC is performed, preventing them from being considered as completely uncorrelated.
%

Consequently, similar to the result presented in~\cite{virgillito2022spatially}, we observe that this coherency decreases as more chromatic dispersion is accumulated between the correlated spans.
As an example, the XCI generated at the 3rd fiber span has an inherent XCI generation, given by $\sigma_3^2$, but also has contributors due to the correlation with the 1st and 2nd spans:
\begin{align}
    \Delta P_{\mathrm{XCI,1}} &= \sigma_1^2\;, \\
    \Delta P_{\mathrm{XCI,2}} &= \sigma_2^2 + C_{21} \sigma_2\sigma_1\;, \\
    \Delta P_{\mathrm{XCI,3}} &= \sigma_3^2 + C_{31} \sigma_3\sigma_1 + C_{32} \sigma_3\sigma_2\;.
\end{align}

The $C_{ij}$ value was estimated by SSFM simulations for every scenario, following Eq~.\ref{eq:cij}, making use of the intrinsic terms given by the green lines in Fig.\ref{fig:pump2_results}.
We have evaluated only the XCI values of the periodic case starting from OLS2 (blue curve of Fig.\ref{fig:pump2_results}), in order to observe the fundamental behavior of the memory-inducing effect, without including the change in gradient when passing from OLS1 to OLS2.
This change in gradient is partially dependent upon the difference in parameters of the two OLSs, and requires a separate in-depth characterization that is beyond the scope of this work.
We remark that the first 10 spans of OLS1 correspond to the first 10 spans of OLS2 with swapped $D_1$ and $D_2$ values. 
This means that evaluating $C_{ij}$ values for OLS2 quantifies the behavior of all investigated cases.

In order to take into account the system parameters of the network segment, we define a variable, $\theta_{\mathrm{span}}$, which quantifies the amount of residual dispersion introduced at each fiber span. This parameter, likewise defined within~\cite{virgillito2022spatially}, is given by:
\begin{equation}
    \theta_{\mathrm{span}}(i,j)=R_s^2\pi\left|\sum_{k=j}^{i-1}\left(\beta_{2,k} L_s+\beta_{\mathrm{DCU},k}\right)\right|\;,
\end{equation}
where $\beta_{2,k}$ is the dispersion coefficient of the kth fiber span, and $\beta_{\mathrm{DCU}}$ is the dispersion introduced by the DCU expressed in terms of frequency.
\begin{figure}[b]
    \captionsetup{singlelinecheck=false, font=small, justification=raggedright}
    \centering
    \begin{subfigure}[t]{0.49\linewidth}
        \includegraphics[width=1\linewidth]{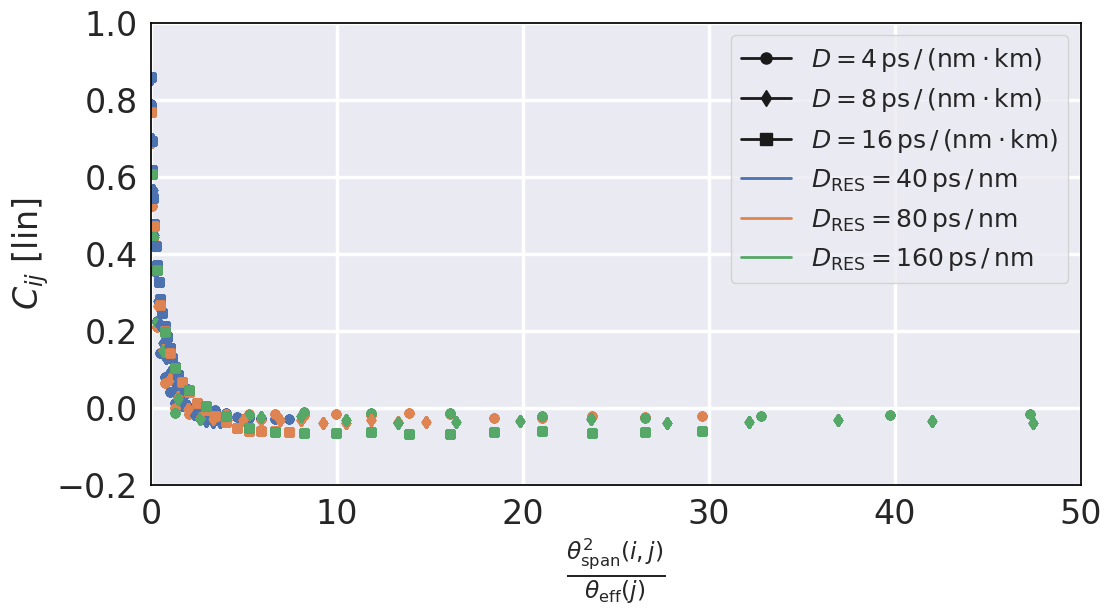}
        \footnotesize
        \vspace{-10mm}
        \caption{}
        \label{fig:cijs_scatter_rs32_df37_pumps02_big}
    \end{subfigure}
    \begin{subfigure}[t]{0.49\linewidth}
        \includegraphics[width=1\linewidth]{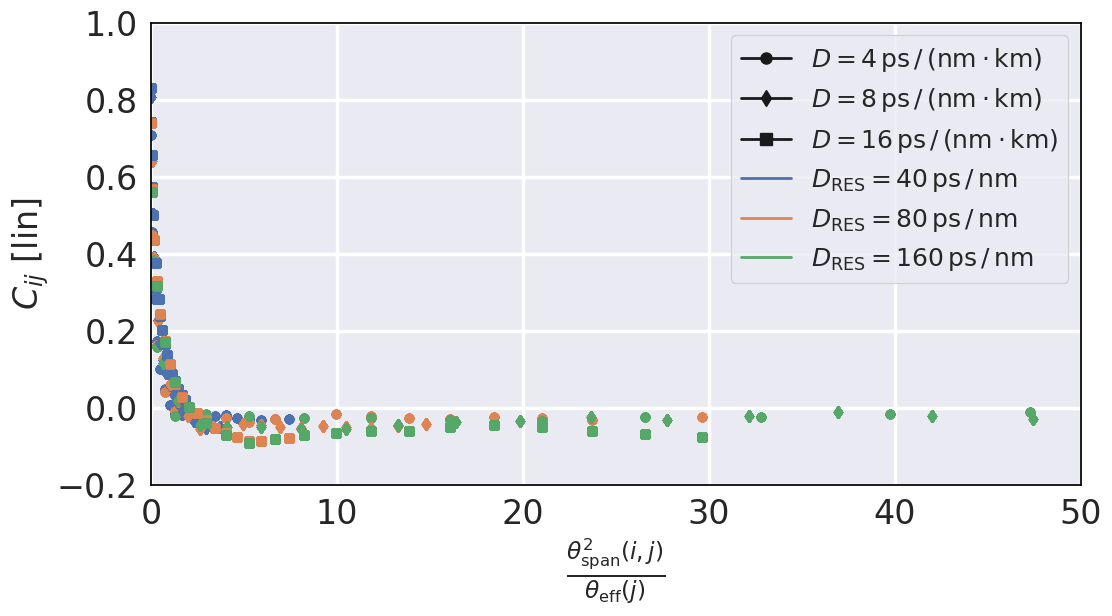}
        \footnotesize
        \vspace{-10mm}
        \caption{}
        \label{fig:cijs_scatter_rs32_df37_pumps04_big}
    \end{subfigure}
    \begin{subfigure}[t]{0.49\linewidth}
        \includegraphics[width=1\linewidth]{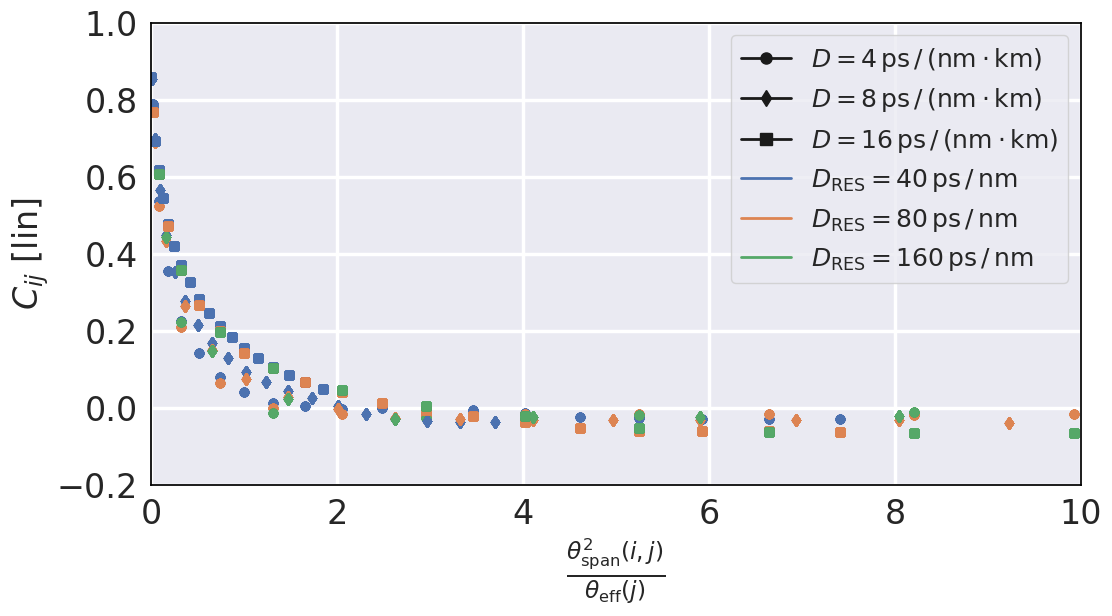}
        \footnotesize
        \vspace{-10mm}
        \caption{}
        \label{fig:cijs_scatter_rs32_df37_pumps02_small}
    \end{subfigure}
    \begin{subfigure}[t]{0.49\linewidth}
        \includegraphics[width=1\linewidth]{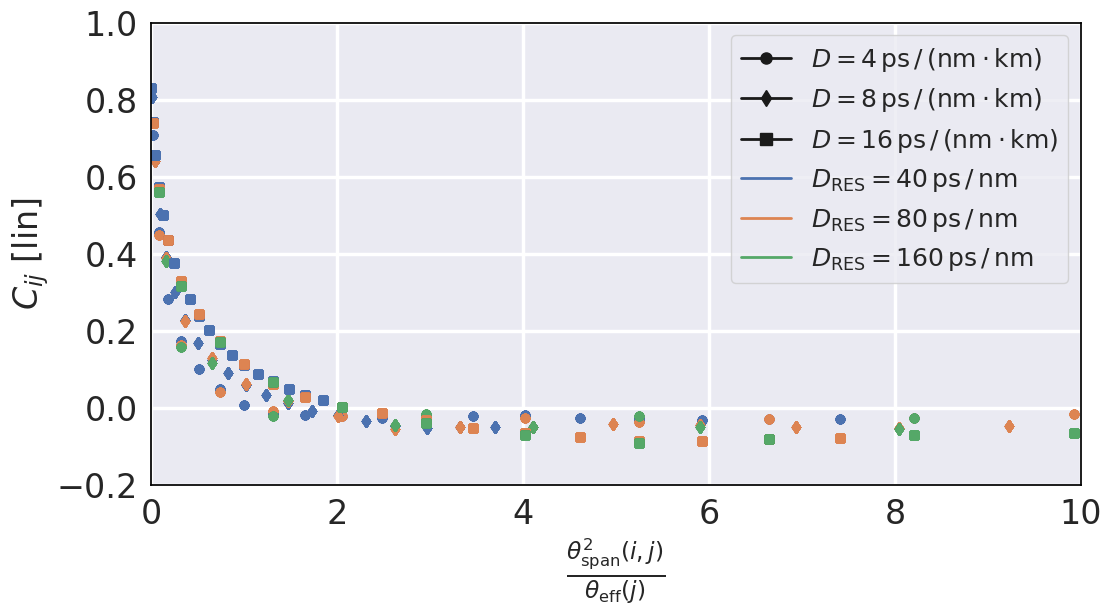}
        \footnotesize
        \vspace{-10mm}
        \caption{}
        \label{fig:cijs_scatter_rs32_df37_pumps04_small}
    \end{subfigure}
    \caption{The plot of the correlation factor, $C_{ij}$, against $\theta_{\mathrm{span}}^2\,/\theta_{\mathrm{eff}}$, which quantifies the behavior of the XCI accumulation depending upon the scaled fiber parameters, for an interfering pump located (a): 150\,GHz and (b): 300\,GHz away. Three different residual dispersion values are shown; $D_{\mathrm{RES}}=40$, 80 and 160\,ps\,/\,nm are given by the blue, orange and green dots, respectively. The dense regions of interest in (a) and (b) are shown more closely in (c) and (d), respectively.}
    \label{fig:cijs_pump2}
\end{figure}
This parameter may then be normalized with respect to another variable, $\theta_{\mathrm{eff}}$, which takes into account all inherent system parameters that the XCI accumulation was observed to scale with in Fig.~\ref{fig:pump2_results}:
\begin{equation}
    \theta_{\mathrm{eff}} = \pi R_s^2 \beta_{2} L_{\mathrm{eff}}\;.
\end{equation}

The calculated $C_{ij}$ values are plotted against $\theta_{\mathrm{span}}(i,j)$\,/\,$\theta_{\mathrm{eff}}$ in Fig.~\ref{fig:cijs_pump2}, for all three considered $D$ values; Fig.~\ref{fig:cijs_scatter_rs32_df37_pumps02_big} and~\ref{fig:cijs_scatter_rs32_df37_pumps04_big} show the results for an interfering pump 150 and 300\,GHz away from the CUT, respectively.
A closer view of these results are shown in the same figure within~\ref{fig:cijs_scatter_rs32_df37_pumps02_small} and~\ref{fig:cijs_scatter_rs32_df37_pumps04_small}, respectively, highlighting the region around the origin, where the correlation most significantly affects the values of $\Delta P_{XCI}$.
Firstly, it is visible in these figures that the behaviour for all three $D_{\mathrm{RES}}$ cases follows the same curve, albeit with a small difference in gradient.
This implies that the correlation between two given spans $i$ and $j$ decreases with an inverse proportionality to the total $D_{\mathrm{RES}}$, from span $j$ to $i-1$, identically to the observed behaviors within Fig.~\ref{fig:pump2_results}.
We observe that $C_{ij}$ decreases proportionally to $\theta_{\mathrm{span}}(i,j)$\,/\,$\theta_{\mathrm{eff}}$.
Importantly, all $C_{ij}$ values are below zero after a given $\theta_{\mathrm{span}}(i,j)$\,/\,$\theta_{\mathrm{eff}}$ value, which means that the coherency induced by the residual dispersion will decay to zero.
Subsequently, given a long enough distance, the XCI will accumulate towards the asymptotic value defined by the parameters of the fibers within OLS1, with this happening faster for higher $D_{\mathrm{RES}}$ values.

Changing the frequency distance between the CUT and interfering channel provides only small differences to the distributions of the data, without any major alterations to the overall behavior.
This implies that, for each CUT and interfering channel pair, there exists a curve which characterizes the relationship between the coherent effect induced by the residual dispersion and the system parameters, enabling this approach to be applied to any CUT and interfering channel pair within a disaggregated network framework.
As a final remark, we note that the behavior of the XCI accumulation bears significant similarity to the behavior of the SCI accumulation~\cite{d2020quality,virgillito2022spatially}, suggesting that these parameters provide an intrinsic characterization of coherent NLI contributions, for both SCI and XCI, in both uncompensated and dispersion-managed scenarios.

\section{Conclusion}
\label{sec:conclusion}
Within this work we have presented an analysis of a variety of pump-and-probe transmission configurations for a segment within a disaggregated network, performing coherent transmission is over dispersion-managed links in order to analyse the generation of XCI.
We observe that, as the amount of residual dispersion present within the link after correction by a DCU increases, so does the discrepancy between the expected XCI, given by the IGN model, and the actual value attained.
This discrepancy evolves towards an asymptotic upper bound, which has a definite value that depends upon the chromatic dispersion of the first fibers within the segment under investigation, and scales according to the parameters of this fiber and the baud rate of the signal being transmitted.
This upper bound can be calculated and therefore used to quantify the amount of additional XCI generated during propagation of coherent signals through dispersion-managed links with low residual dispersion values, ensuring that accurate QoT predictions are still possible in the case where IMDD systems are undergoing modernization.
\section*{Funding}
\begin{wrapfigure}{o}{0.2\linewidth}
    \vspace{-4mm}
    \centering
    \includegraphics[width=0.6\linewidth]{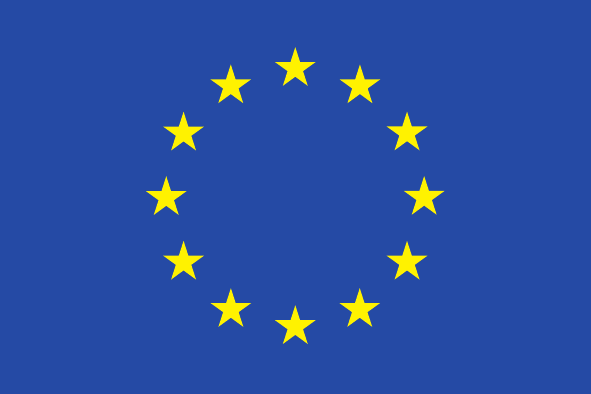}
\end{wrapfigure}
This project has received funding from the European Union’s Horizon 2020 research and innovation program under the Marie Skłodowska-Curie grant agreement 814276.
\section*{Disclosures}
The authors declare no conflicts of interest.
\section*{Data Availability}
Data underlying the results presented in this paper are not publicly available at this time but may be obtained from the authors upon reasonable request.
\bibliography{bibliography}
\end{document}